\documentclass[structabstract]{aa}  
\usepackage{graphicx,natbib}
\usepackage{txfonts}

\begin{document}
   \title{Variation of the extinction law in the \object{Trifid} nebula}

   \author{L. Cambr\'esy\inst{1}
          \and
		J. Rho\inst{2,4}
          \and
		D. J. Marshall\inst{3}
          \and
		W. T. Reach\inst{2,4}
          }

   \institute{Observatoire astronomique de Strasbourg, Universit\'e de
		Strasbourg, CNRS, UMR 7550, 11 rue de l'Universit\'e,
		F-67000 Strasbourg, France\\
              \email{cambresy@astro.unistra.fr}
	\and
              Infrared Processing and Analysis Center, California Institute
	      of Technology, Pasadena, CA 91125, USA
	\and
		Universit\'e de Toulouse, UPS, CESR, 9 avenue du colonel Roche,
		31028 Toulouse Cedex 4, France
              \email{Douglas.Marshall@cesr.fr}
	\and
	      Stratospheric Observatory for Infrared Astronomy, 
	      Universities Space Research Association, 
	      Mail Stop 211-3, Moffett Field, CA 94035, USA
	      \email{jrho@sofia.usra.edu, wreach@sofia.usra.edu}
             }

   \date{Received ; accepted }

 
  \abstract
  {
  In the past few years, the extinction law has been measured in the
  infrared wavelengths for various molecular clouds and different laws have
  been obtained.
  } 
   {
  In this paper we seek variations of the extinction law within the
  \object{Trifid} nebula region. Such variations would demonstrate a local
  dust evolution linked to variation of the environment parameters such as
  the density or the interstellar radiation field. 
   }
   {
  The extinction values, $A_{\lambda}/A_V$, are obtained using the 2MASS,
  UKIDSS and Spitzer/GLIMPSE surveys. The technique is to inter-calibrate 
  color-excess maps from different wavelengths to derive the extinction law
  and to map the extinction in the Trifid region.
   }
   {
  We measured the extinction law at 3.6, 4.5, and 5.8 $\mu$m and we found
  a transition at $A_V\approx20$~mag. Below this threshold the
  extinction law is as expected from models for $R_V=5.5$ whereas above
  20~mag of visual extinction, it is flatter.Using these results the
  color-excess
  maps are converted into a composite extinction map of the Trifid nebula
  at a spatial resolution of 1~arcmin. A tridimensional analysis along
  the line-of-sight allowed us to estimate a distance of $2.7\pm0.5$~kpc
  for the Trifid. The comparison of the extinction with the 1.25~mm emission
  suggests the millimeter emissivity is enhanced in the dense condensations
  of the cloud.
   }
   {
   Our results suggest a dust transition at large extinction which has not
   been reported so far and dust emissivity variations. 
   }

   \keywords{ISM: dust, extinction --
               ISM: individual object: Trifid nebula --
	       Infrared: ISM
               }

   \maketitle
%

\section{Introduction}
During the last three decades, dust evolution in the interstellar medium
has been studied mostly through the grain emission in the far-infrared and
then the sub-millimeter wavelengths. The analysis of the surface brightnesses
at 60 and 100~$\mu$m observed by IRAS lead to distinguish a small and a big
grain component where the small grains are the major contributor at
60~$\mu$m \citep{BBD88,LCP91}. The transition from small to big grains
occurs at a typical visual extinction of 1~mag. 
More recently a new transition was discovered in the dust composition with
the formation of fluffy grains. The analysis of PRONAOS sub-millimeter
observations by \citet{BAR+99} yielded to the discovery of a population of
enhanced-emissivity dust grains which have a lower equilibrium temperature.
This emissivity enhancement in the far-infrared by a factor of about 3
\citep{CBLS01} requires the existence of composite grains formed by the
coagulation of small grains. A large scale study on the whole Galactic
anticenter hemisphere showed this transition also occurs at low extinction,
$A_V\approx 1$~mag \citep{CJB05}. It might be closely related to the recent
discovery made by \citet{MSB+10a} with Herschel data where this same density
threshold seems to indicate the transition where diffuse clouds start to
collapse into filaments.

With the wealth of available deep data in the near and mid-infrared it is now
possible to investigate the interstellar grain properties at shorter
wavelengths. In this spectral domain a lot of progresses has been made
in the past few years through the study of the extinction law.
Dense clouds can be probed on large scales and for various
astrophysical environments. Recent studies with Spitzer data
\citep{IMB+05,FPM+07,RLMA07, CMLE09} focused on different clouds and
obtained different extinction laws. We propose in this work to look for
possible variations of the extinction curve within a single cloud.
We choose an active star-forming region which includes the famous
HII region M20, the Trifid nebula. Its location in the Galactic
plane at only 7~degrees from the Galactic center offers a high stellar
density which is ideal to probe the extinction distribution. The distance
of this young nebula ($10^5$ years) is still debated. The observation of
the central O7 star responsible for the photoionization, \object{HD 164492},
yields a distance from 1.67~kpc \citep{LCO85} to 2.8~kpc \citep{KML99}.
This active region has its first generation of massive stars interacting
with the interstellar medium and is very likely triggering a second generation
of star formation \citep{CLC+98}. The state-of-the-art for the Trifid
star-forming region is presented in a review paper by \citet{RLRC08}.

The extinction law is generally estimated in molecular clouds as a whole
and the discrepancies between the different results are interpreted as
variations of the dust optical properties from cloud-to-cloud due to the
environment. Obviously variations should also be observed within molecular
clouds. As the density increases from the envelope to the core of a cloud,
the dust properties, hence the extinction law, should be affected.
An attempt to detect such a variation was proposed by \citet{RLMA07} in
their study of a dark cloud core, B59, up to $A_V\approx59$~mag but they
did not succeed. \citet{OO10} used five different methods to derive the
extinction law in B335. The scatter in their result was consistent with the
uncertainties so they could not conclude on any evolution of the
extinction law within the cloud.
\citet{CMLE09} managed to find variations by averaging
their result from several clouds and they showed evidence for grain
growth at $A_{K_s}>1$~mag ($A_V>9$). A global galactocentric variation
of the extinction law has also been reported by \citet{ZMI+09} in
a large scale study that excluded molecular clouds.

The present study focuses on the variations of the extinction law within the
molecular cloud associated with the Trifid nebula. Our analysis relies on
the 2MASS \citep{SCS+06}, UKIDSS \citep{LWA+07} and GLIMPSE II \citep{CBM+09}
surveys. Special attention is given to the data filtering to limit the
possible biases as explained in Sect.~\ref{s.data}.
Our original method to derive the extinction law from 3.6 to 5.8~$\mu$m
is based on the color mapping rather than the classical analysis
of the source list distribution in a color-color diagram. The details and
the advantages of this technique are presented in Sect.~\ref{s.extlaw}
while Sect.~\ref{s.extmap} describes the final extinction map making and
a tridimensional analysis of the Trifid line-of-sight which allows us to
disentangle the complexity of this direction. In addition we studied the
dust emissivity at 1.25~mm by comparing the masses derived from the dust
continuum emission and the dust absorption.
We conclude in Sect.~\ref{s.conclusion}.

\section{Data preparation}\label{s.data}
\subsection{Datasets}
Our study covers a large field of $1.6^\circ \times 1.4^\circ$.
The 2MASS near-infrared data provides $JHK_s$ fluxes for the whole sky. The
high stellar density in the Trifid direction induces a loss of sensitivity
as 2MASS sources with less than 5\arcsec\ separation are affected by confusion.
The degradation reaches almost 1.5 mag in each filter, with completeness
level of only 12.8 mag for the $K_s$ band for instance. 

Deeper $JHK_s$ fluxes are obtained from the UKIDSS 6th data release
(UKIDSSDR6plus) of the Galactic Plane survey. UKIDSS uses the UKIRT
Wide Field Camera \citep[WFCAM;][]{CAA+07}. The photometric system is
described in \citet{HWLH06}, and the calibration is described in
\citet{HIHW09}. The pipeline processing and science archive are
described in \citet{HCC+08}.

Longer wavelengths from Spitzer/IRAC are publicly available through the
GLIMPSE II legacy survey. We used the {\em high reliability
point source Catalog} (v2.1) which includes data from epochs 1 and 2.

The merged catalog between 2MASS, UKIDSS and GLIMPSE tables results from
a 1\arcsec\ cone search. Completeness limit magnitudes for all filters are
presented in Table \ref{t.compl}.
For bright stars, the 2MASS photometry is chosen over the UKIDSS photometry
which is affected by saturation.
Although the UKIDSS photometry is directly calibrated on 2MASS we found a
systematic difference between the two systems: 
$$
\begin{array}{lll}
J\mbox{(2MASS)}   &= J\mbox{(UKIDSS)}   & + 0.084\\
H\mbox{(2MASS)}   &= H\mbox{(UKIDSS)}   & - 0.074\\
K_s\mbox{(2MASS)} &= K_s\mbox{(UKIDSS)} & - 0.003
\end{array}
$$
We corrected the UKIDSS photometry to match the 2MASS system.
The UKIDSS astrometry has also been corrected to match
the GLIMPSE positions by adding 0.12 and 0.07~arcsec to the right
ascension and declination, respectively.
In the following, only magnitude measurements with an uncertainty smaller
than 0.15 mag will be considered.  It yields about $2\times10^5$ sources
at $K_s$, $6\times10^5$ at 3.6 $\mu$m in the range 
$\mbox{R.A.} \in [269.50,271.29]$ and $\mbox{Dec.} \in [-23.6, -22.2]$.

\begin{table}
\caption{Completeness limit magnitudes. For UKIDSS the luminosity function
slopes are not well defined, so the peak values of the function for
sources with an error smaller than 0.15~mag are displayed.}\label{t.compl}
\begin{tabular}{ccccc}
\hline\hline
        & 2MASS & UKIDSS & UKIDSS(peak)& GLIMPSE \\
\hline
$J$     & 14.4  & 16.2   & 17.8        &         \\
$H$     & 13.5  & 15.0   & 16.7        &         \\
$K_s$   & 12.8  & 14.2   & 16.0        &         \\
$[3.6]$ &       &        &             & 13.5    \\
$[4.5]$ &       &        &             & 13.4    \\
$[5.8]$ &       &        &             & 12.2    \\
$[8.0]$ &       &        &             & 11.4    \\
\hline
\end{tabular}
\end{table}

\subsection{Source selection criteria}\label{s.selection}
The analysis of the stellar color excess to investigate the cloud properties 
requires all stars to be background stars. The contamination by foreground
objects cannot be neglected for the Trifid nebula located between 1.7 and
2.8~kpc probably in the Scutum Arm, behind the Sagittarius Arm.
The procedure to remove foreground stars is statistical for two reasons:
1) the number of detected sources is close to $6\times10^5$, and 2) there is a
degeneracy between intrinsically red foreground objects and blue background
sources. A color criteria would therefore be inefficient. It would actually
lead to a source spatial distribution correlated with the dust distribution
which is not consistent with the supposedly foreground nature of the stars.
The individual identification is generally impossible, except in the
high extinction region where the degeneracy can be raised. 
Keeping these constraints in mind, we choose a two steps strategy,
first purely statistical and then individual and manual.

The statistical step consists in sampling the source catalog into a data
cube with two spatial dimensions and a third dimension being a color axis.
Only one star per box is allowed. 
Basically, stars are sampled by a 4~arcmin grid step according to their
celestial coordinates and the third axis allows to color-sort the stars
within a same 16~arcmin$^2$ pixel so that there is only a single star per
3D-box. The first slice of the cube represents a map where each pixel
contains the color of the bluest object of the stack. The following slices
contain redder and redder color values.
The completeness limit of UKIDSS being not well defined we use only 2MASS
and GLIMPSE for this statistical procedure.
The variation of the stellar density makes the number of stars per pixel,
and therefore the size of the stack, range from 85 to 387. 
For a 16~arcmin$^2$ pixel, it means the source density varies from
$19\times10^3$ to $87\times10^3$~deg$^{-2}$.
Once the cube is built, each plane from the bluest to the reddest
is compared to a raw color map of the field, i.e. a non-calibrated
extinction map. A correlation with the color map starts at the
38$^{\rm th}$ slice, meaning the distribution of the $\sim$38
bluest stars in all the 16~arcmin$^2$ pixels is not correlated
with the dust distribution.
They must be considered as foreground objects and the first 38 stars of each
pixel, i.e. 8600 stars deg$^{-2}$, are thus removed from the original catalog.
It represents from 10\% to 45\% of the sources detected at $K_s$ and $[3.6]$
depending on the position, i.e. the local source density.

At this point, only a fraction of the foreground stars is removed. A second
step is required to filter the catalog in the regions where the contamination
is critical because the foreground objects may even be the dominant population.
Fortunately, in these regions the color degeneracy is raised by the strong
reddening from the dust grains. To define these regions we build a [3.6]-[4.5]
color map following the adaptive method described in Sect.~\ref{s.mapmethod}.
Then we select the pixels where [3.6]-[4.5]$>$0.35~mag ($A_V\approx30$ mag)
which yields a 25~arcmin$^2$ surface area. Such a high threshold guaranty
the degeneracy with the intrinsic star color is broken.
For this manual operation the restriction on UKIDSS data can be waived and
$\sim$$10^3$ sources are detected in the 25~arcmin area at both $K_s$ and
3.6 $\mu$m with error smaller than 0.15 mag.
The spectral energy distribution from 1.25~$\mu$m to 8~$\mu$m is examined
for these sources. For each foreground star candidate, a color-composite
$JHK_s$ image is built to compare the color of the candidate itself with
the color of the surrounding objects.
This visual inspection allows to make the final classification of the
source since a foreground star appears bluer than the surrounding objects.
Among the $\sim$$10^3$ sources detected at $K_s$ and 3.6 $\mu$m, 20\% are
classified as foreground stars. It corresponds to a total foreground star
density of $\sim$$37\times10^3$~deg$^{-2}$.
It worth noting that for sources together detected at $J$, $H$ and $K_s$
instead, the contamination level in the 25~arcmin$^2$ reaches 55\%. This is
because many background stars are lost when the $J$ detection is required.

Beside the foreground stars, there is a contamination by the Young
Stellar Objects (YSOs) that are formed in the molecular cloud associated with
the Trifid nebula. Several studies, at various wavelengths, provide us with
a source list of YSOs that we removed from our catalog. The most important
contribution is from \citet{RRLF06} who analyzed the infrared colors from
Spitzer IRAC and MIPS data. It yielded the identification of 161 YSOs 
(37 class I/O, 13 hot excess, 111 class II). Previous studies of the
young star population in the Trifid nebula used 2MASS, ASCA, ROSAT
\citep{RCCR01} and Chandra \citep{RRC+04}. 

Finally, another possible contamination would be extragalactic sources.
This effect is negligible based on the location of the Trifid nebula in the
Galactic plane and close to the Galactic Center direction.

\section{Extinction law}\label{s.extlaw}
\begin{figure}
\includegraphics[width=8.8cm]{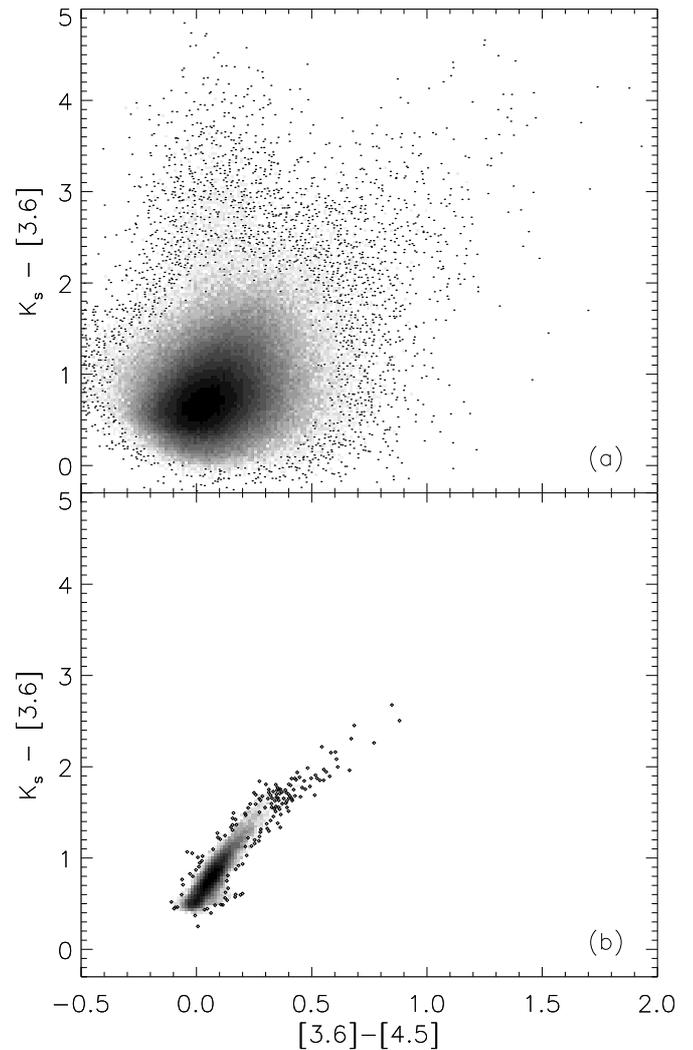}
\caption{{\em (a)} Color-color diagram for stars with uncertainty smaller
than 0.15 mag in the 3 bands. {\em (b)} Color-color diagram obtained
using color maps instead of the source list. The exact same stars are used
for both plots. The gray scale represents the density of the data points
in the plot and the dots show the individual data points for low density.} 
 \label{f.colcol1}
\end{figure}
The general idea of the method to derive the extinction law is trivial.
It requires the knowledge of the relative extinction in two wavelengths,
$A_{\lambda1}/A_V$ $A_{\lambda2}/A_V$, to evaluate the extinction at
a third wavelength by measuring the slope of the star color distribution
within a color-color plot 
$m_{\lambda1}-m_{\lambda2} / m_{\lambda2}-m_{\lambda3}$. The color-color
diagram slope being equal to the color-excess ratio 
$E_{\lambda1-\lambda2} / E_{\lambda2-\lambda3}$ the value of
$A_{\lambda3}/A_V$ is inferred.
Practically, this procedure is not necessarily that simple because of the
scatter in the color-color diagram. The selection of a sub-sample as clean as 
possible is strongly recommended. Fig.~\ref{f.colcol1}a is restricted to sources
with an uncertainty smaller than 0.15~mag at $K_s$, 3.6 and 4.5~$\mu$m
although a large scatter is still present in the plot. In particular, the
vertical elongation for $[3.6]-[4.5]\approx0$ and $K_S-[3.6]>2$ is not real.
It is due to a severe underestimation of the UKIDSS $K_s$ uncertainty for
some sources. Fortunately, the method presented in the next section makes
their contribution negligible because of their random spatial distribution. 

Fig.~\ref{f.colcol1} and several others in this paper represent plots as density
maps for clarity reason. Dark means high density. Unless specified otherwise
the overlaid symbols simply help to visualize the lowest density pixels that
are hardly visible.

\subsection{Color map making}
\subsubsection{Method}\label{s.mapmethod}
Although a slope could be measured in Fig.~\ref{f.colcol1}a, the scatter is
large. The Trifid line-of-sight crosses various interstellar medium
environments and all kind of stellar populations. This diversity, the high
stellar density, and the relatively large distance of M20 make our goals
more difficult to reach than for nearby molecular clouds such as
Chamaeleon, Taurus or Rho Ophiuchus.

We propose a method based on the color maps that spectacularly
improves the accuracy of the analysis. Fig. \ref{f.colcol1}b was built using
the exact same sources as Fig.~\ref{f.colcol1}a and both figures have the same
axis, same scale. While each point in Fig.~\ref{f.colcol1}a corresponds to a
single source, points in Fig.~\ref{f.colcol1}b actually represent individual
pixels taken from two color maps, one for each axis. Using the 
color allows us to keep the same axis as in Fig.~\ref{f.colcol1}a
but it is obviously equivalent to an extinction map assuming an extinction
law. The impressive difference between the two figures highlights the
superiority of using maps rather than a source list directly.
Using maps permits to average the color of several individual nearby
sources, and thus to reduce the scatter.
Different methods are available to build an extinction map.
Star count methods are sometimes used on large scale
and for moderate extinction \citep{Cam99d,DUK+05}. This approach is
definitely not adapted to a highly obscured distant cloud in the
Galactic plane. Too sensitive to the foreground star contamination
the map would be noisy and seriously biased.
Color-excess methods are better adapted. The first statistical method
to map the extinction from color excess, called NICE, was proposed by
\citet{LLCB94}. Improvement were proposed and called NICER \citep{LA01} 
and then NICEST \citep{Lom09}. A variant technique from \citet{CBJC02}
proposes to replace the arithmetic mean color by the median color in cells
and to fix the number of sources within a cell rather than its size. The
latter step makes the method adaptive since the spatial resolution is always
optimal, adjusted to the local density. A final convolution can be applied if
a fixed spatial resolution is wished. This convolution minimizes a bias that
would result in an underestimation of the extinction, compared to the usage
of larger
cells already at the final resolution size. Moreover the median is a
powerful tool to limit the effect of the residual contamination by foreground
stars and young stellar objects as detailed in \citet{CBJC02}.

All the color maps built for this work use the median color and
follow the adaptive cell scheme with a fixed number of 3 stars per cell.
The variable cell spatial resolution being $r_i$, a convolution by a Gaussian
kernel of $\sqrt{1-r_i^2}$~arcmin is applied to reach a final resolution of
1~arcmin over the whole map.

\subsubsection{Selection bias}
\begin{figure}
\includegraphics[width=8.8cm]{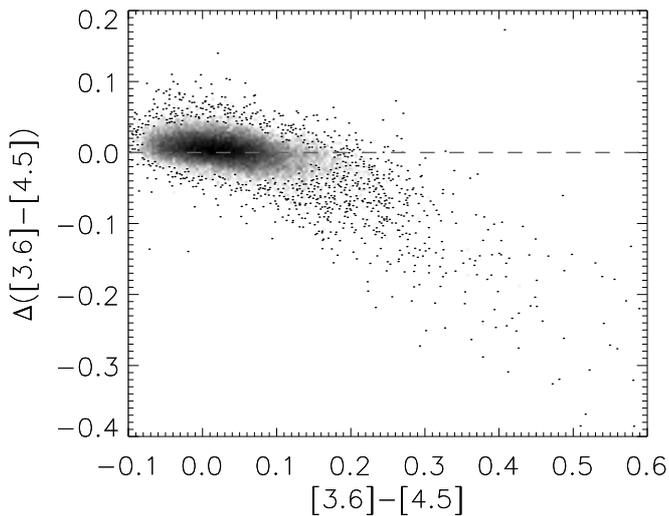}
\caption{$[3.6]-[4.5]$ color difference from sources detected
simultaneously in two different sets of 3 filters, $K_s({\rm 2MASS})/
[3.6]/[4.5]$ and $[3.6]/[4.5]/[5.8]$, versus color.}
 \label{f.bias}
\end{figure}
Once a clean star sample is produced, i.e. filtered from foreground stars,
YSOs and noisy photometry sources, the color maps are
still subject to a selection bias. While it is mandatory to keep sources
detected in all 3 bands to plot Fig.~\ref{f.colcol1}a, the
Fig.~\ref{f.colcol1}b needs two color maps that can be built
independently with only a 2 band detection criteria for each of them.
Fig.~\ref{f.bias} illustrates the color bias for $[3.6]-[4.5]$ when it is
obtained from two different sets of 3 band detection criteria, 
$K_s/[3.6]/[4.5]$ or $[3.6]/[4.5]/[5.8]$. To exaggerate the effect $K_s$ is
taken from 2MASS which is not as deep as UKIDSS.
The $[3.6]-[4.5]$ average color is affected by the $K_s$ detection criterion
which makes the color bluer, and by the [5.8] detection criterion which tends
to select redder objects. It is not a simple offset as shown in
Fig.~\ref{f.bias} because this effect is more important for redder color.
In terms of extinction, taking sources detected in the 3 bluer wavelengths
yields an underestimation of the extinction. Using deeper $K_s$ data would
obviously help since more sources detected at [3.6] and [4.5] would also be
seen at $K_s$. However the effect cannot be avoided at large extinction
because the $K_s$ band suffer from twice the extinction than at [3.6].

It is therefore critical to compare only maps that are built with the
exact same sample of stars. It implies the  $[3.6]-[4.5]$ color map
needs to be built from a source list which contains stars detected in
the 3 bands $K_s$, [3.6] and [4.5] if it is compared to a $K_s-[3.6]$ map,
or  from a source list which contains stars detected at [3.6], [4.5] and
[5.8] if it is compared to a $[4.5]-[5.8]$ map.  

A total of 6 color maps are built for our study: $H-K_s$, $K_s-[3.6]$,
$[3.6]-[4.5]$, $[4.5]-[5.8]$. Each map is obtained with sources detected in 3
bands at the same time. There are 2 versions for the $K_s-[3.6]$ map and the
$[3.6]-[4.5]$ map because the third band can either be the bluer or the redder
one.

\subsection{Results}
\begin{figure}
\includegraphics[width=8.8cm]{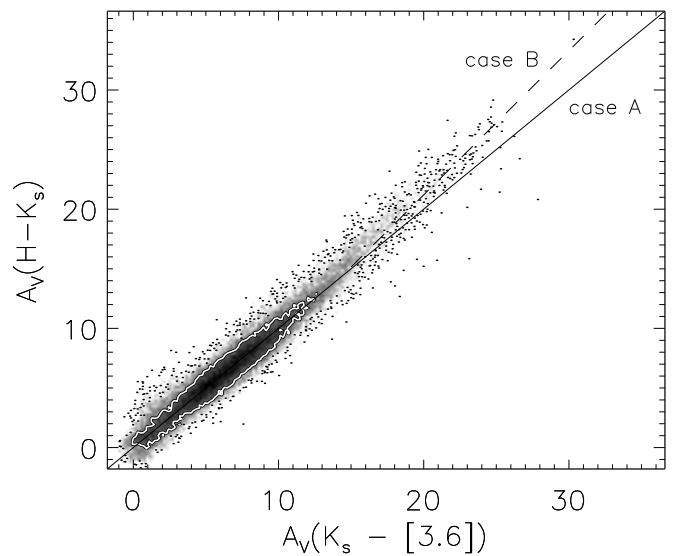}
\caption{Extinction from $H-K_s$ color excess versus extinction from
$K_s-[3.6]$ color excess. By adjusting $K_s-[3.6]$ to obtain a one-to-one
regression line (solid line) we find $A_{[3.6]}/A_{K_s}=0.611$. The white
iso-density contour delineates the region  used to compute de regression line.
The dashed line is the regression line for $A_V>20$~mag, it indicates the
deviation at large extinction. Case A and B refer to Table~\ref{t.res1}.} 
\label{f.HK-KI1}
\end{figure}
Although the extinction law has been for long considered as universal in the
infrared it is now admitted there are variations at wavelength longer than
2~$\mu$m. The spectral range that is pseudo-universal is actually restricted to
1-2~$\mu$m where the extinction is independent or very little dependent of the
environment, i.e. dust temperature and density. Consequently,
$A_H/A_V=0.175$ and $A_{K_s}/A_V=0.112$ \citep{RL85} are assumed constant.  
It corresponds to $\beta=1.7$ in the expression
$A_\lambda \propto \lambda^{-\beta}$ that is often used to describe the
near-infrared extinction law. Values from 1.6 to 1.8 being reasonable, the
effect of such a systematic uncertainty is discussed further in this section.
The $H-K_s$ map can be converted into an extinction map using
$$A_V = \frac{(H-K_s)-(H-K_s)_0}{A_H/A_V - A_{K_s}/A_V} = 15.87 \; E_{H-K_s}$$
The extinction law is normalized to $A_V$ mainly for historical reason but it
is more convenient to normalize to $K_s$. The relation between the extinction
and the color excess becomes $A_{K_s} = 1.78 \; E_{H-K_s}$.
The reference color $(H-K_s)_0=0.54$~mag is estimated from the area
with the smallest values within the map, about 50~arcmin south-west from the
densest core of the cloud. This is much redder than the intrinsic stellar color
because it includes the diffuse extinction along the line-of-sight.
The reference color is relevant to provide an absolute extinction for the
Trifid alone and it is discussed in the Sect.~\ref{s.3d}. It has absolutely
no effect on the extinction law estimation nor on the relative extinction
comparison from different parts of the field.

Fig.~\ref{f.HK-KI1} represents $A_V$ from $H-K_s$ versus $A_V$ from
$K_s-[3.6]$. Since $A_V=A_{K_s}/0.112$ is assumed, and by definition
$A_{K_s}= E_{K_s-[3.6]} / (1 - A_{[3.6]/A_{K_s}})$, the only free
parameter to adjust to obtain a slope of 1 is the value of $A_{[3.6]}/A_{K_s}$. 
After the adjustment, it yields the value of $A_{[3.6]}/A_{K_s}=0.611$.
Fig.~\ref{f.HK-KI1} is presented as a density map for clarity and also
because the density information is taken into account in the fitting
process to exclude the low density pixels. Basically the minimum density
level is fixed to 20\% of the maximum density in the map.
The color-coding used in the density plots is in log-scale so the
20\% always corresponds to the dark region. The white iso-density contour
delineates the region selected to compute the regression line in
Fig.\ref{f.HK-KI1}. Any outlier is therefore fully eliminated
with this technique.

\begin{figure*}
\includegraphics[width=18cm]{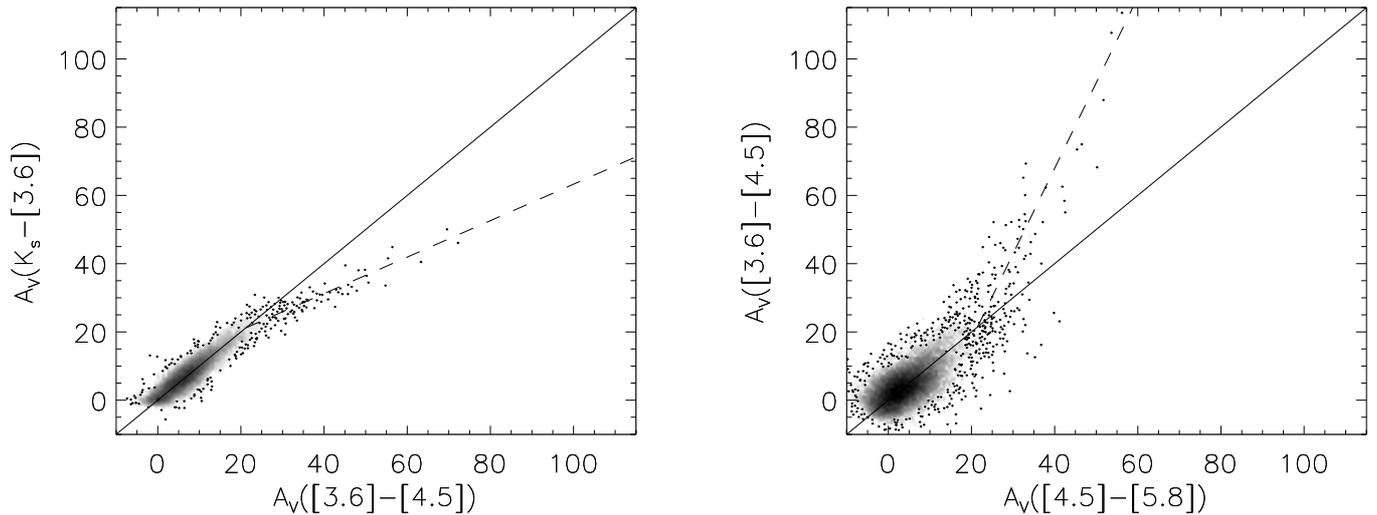}
\caption{Each plot represents the extinction correlation from two different
color excesses. The two parameters $A_{[4.5]}/A_{K_s}=0.500$ (left plot)
and $A_{[5.8]}/A_{K_s}=0.398$ (right plot) are adjusted to obtain a
one-to-one regression line (solid line).
The dashed lines trace the deviation at large extinction, the
corresponding slopes in these plots are $0.55\pm0.02$ and $2.52\pm0.09$,
respectively.} 
 \label{f.KI1I2I3}
\end{figure*}

Once $A_{[3.6]}/A_{K_s}$ is determined, $A_{[4.5]}/A_{K_s}$ is evaluated with
a plot of $A_V$ from $K_s-[3.6]$ versus $A_V$ from $[3.6]-[4.5]$. A third
similar plot will then provide $A_{[5.8]}/A_{K_s}$ (see Fig.~\ref{f.KI1I2I3}).
The results are $A_{[4.5]}/A_{K_s}=0.500$ and $A_{[5.8]}/A_{K_s}=0.398$. 

Since the pixels falling in the low density regions of the 
$A_V$ vs. $A_V$ plot are ignored the fitting is performed on regions 
where the visual extinction is roughly smaller than 15~mag.
At large extinction a striking deviation of the slopes is observed and
outlined by the dashed lines in Figs.~\ref{f.HK-KI1} and \ref{f.KI1I2I3}.
The changes of slope occur at $A_V\approx20$~mag ($A_{K_s}\approx2.2$)
and suggests a transition phase in the dust grain population.
At such a high column density this result is unexpected.
Contrary to studies in the far-infrared wavelengths which rely on the
dust emission, there is no temperature degeneracy issue since it is based
on stellar color mapping. A drawback of
this technique is its limitation to the near-infrared spectral range.
At wavelengths longer than $\sim$5~$\mu$m the number of detected stars
is too low to have an acceptable spatial sampling of the molecular clouds
while shorter wavelengths are so sensitive to the extinction that the
cloud dense regions become totally opaque.
A direct consequence on this limitation is that no evidence for a slope
change can be seen in a $J-H$ vs. $H-K_s$ diagram because the visual
extinction cannot go deeper than $\sim$20~mag with $J-H$. That prevents
any conclusion on the extinction law variation at shorter wavelengths
than 3.6~$\mu$m.
\begin{table*}
\caption{Color excess ratios and resulting extinction law parameters.
The two rows for $A_V>20$ mag represent different cases for
$A_{[3.6]}/A_{K_s}$: case A propagates the same value as for small
extinction, case B takes the actual measured value.
We assume the power-law index $\beta=1.7$ (i.e. $A_H/A_{K_s}=1.563$) and
results for $\beta=1.6$ and 1.8 are provided for case B. 
The main results from the literature for dense interstellar medium
directions and from a dust model are indicated for comparison.} 
\label{t.res1}
\centering
\begin{tabular}{l c c c c c c c}
\hline\hline
 & $E_{H-K_s}/E_{K_s-[3.6]}$ & $E_{K_s-[3.6]}/E_{[3.6]-[4.5]}$ &
 $E_{[3.6]-[4.5]}/E_{[4.5]-[5.8]}$ & $A_H/A_{K_s}$ & $A_{[3.6]}/A_{K_s}$ & $A_{[4.5]}/A_{K_s}$ &  $A_{[5.8]}/A_{K_s}$ \\
\hline
$A_V<15$ & $1.446\pm0.002$ & $3.508\pm0.011$ & $1.087\pm0.007$ &
1.563 & $0.611\pm0.002$ & $0.500\pm0.003$ & $0.398\pm0.005$ \\ 

$A_V>20$ & & & & & & & \\ 

$\;\;$case A & $1.446\pm0.002$ & $1.915\pm0.018$ & $1.496\pm0.089$ &
1.563 & $0.611\pm0.002$ & $0.408\pm0.003$ & $0.272\pm0.011$ \\ 

$\;\;$case B & $1.749\pm0.003$ & $2.317\pm0.027$ & $1.810\pm0.091$ &
1.563 & $0.678\pm0.006$ & $0.540\pm0.003$ & $0.463\pm0.007$ \\ 

& & & & & & & \\ 

$\;\;$case B, $\beta=1.6$ &  &  &  &
1.520  & $0.703\pm0.005$ & $0.573\pm0.003$ & $0.500\pm0.007$ \\ 

$\;\;$case B, $\beta=1.8$ &  &  &  &
1.601 & $0.656\pm0.006$ & $0.507\pm0.003$ & $0.424\pm0.007$ \\ 

& & & & & & & \\ 
\hline
& & & & & & & \\ 

\multicolumn{4}{l}{Rom\'an-Z\'u\~niga et al. (2007) for Barnard 59} &
1.55\hspace{1ex} & $0.619\pm0.029$ & $0.512\pm0.022$ & $0.459\pm0.024$ \\ 

\multicolumn{4}{l}{Flaherty at al. (2007) for Orion}    &
1.55\hspace{1ex} & $0.636\pm0.003$ & $0.540\pm0.003$ & $0.504\pm0.003$ \\ 

\multicolumn{4}{l}{Chapman et al. (2009) for Ophiuchus, Perseus and Serpens}            &
1.52\hspace{1ex} & $0.640\pm0.030$ & $0.530\pm0.030$ & $0.460\pm0.030$ \\ 

& & & & & & & \\ 
\multicolumn{3}{l}{Weingartner \&Draine (2001) with $R_V=3.1$} & &
& 0.40& 0.25& 0.17\\ 
\multicolumn{3}{l}{Weingartner \&Draine (2001) with $R_V=5.5$} & &
& 0.60& 0.49& 0.40\\ 

\hline
\end{tabular}
\end{table*}
\begin{figure}
\includegraphics[width=8.8cm]{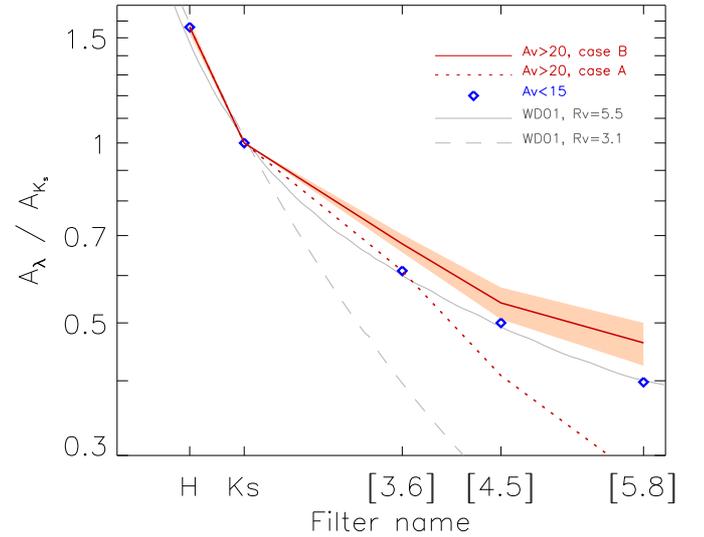}
\caption{Extinction law. The two red lines corresponds to the cases
described in Table~\ref{t.res1} for $A_V>20$~mag. 
The shaded area indicates the systematic error, for case B only,
resulting from $A_H/A_{K_s}$ uncertainty which can range from 1.52
($\beta=1.6$) to 1.60 ($\beta=1.8$). At longer wavelengths than $K_s$,
$\beta=1.6$ gives larger $A_\lambda/A_{K_s}$.
Diamonds are for $A_V<15$~mag. The \citet{WD01} model is plotted in gray
with dashed and plain lines for $R_V=3.1$ and $R_V=5.5$, respectively.
The axis scale is linear for $x$ and logarithmic for $y$.}
 \label{f.extlaw}
\end{figure}
Table~\ref{t.res1} summarizes the results obtained from the slope measured in
both Figs.~\ref{f.HK-KI1} and \ref{f.KI1I2I3}. We consider two different
cases for $A_{[3.6]}/A_{K_s}$ at $A_V>20$~mag:
\begin{itemize}
\item case A: $A_{[3.6]}/A_{K_s}=0.611$ is independent from the extinction,
the value remains identical than at lower extinction (solid line in
Fig.~\ref{f.HK-KI1}).
\item case B: $A_{[3.6]}/A_{K_s}=0.678$ which is the value actually obtained
from the slope of the dashed line in Fig.~\ref{f.HK-KI1}.
\end{itemize}
The statistical errors on the slope estimations are small and so are the
errors on the computed $A_\lambda/A_{K_s}$.
The systematic error on $A_H/A_{K_s}$ is the major source of uncertainty.
We assume a value of 1.563 which corresponds to a power-law index of 
$\beta=1.7$ for the extinction law. 
Our results from Table~\ref{t.res1} are plotted in Fig.~\ref{f.extlaw}.
The shaded area delineates the systematic uncertainty for $\beta$ varying
from 1.6 to 1.8. Using $\beta=1.6$, such as in \citet{CMLE09},
increases $A_\lambda/A_{K_s}$ to the upper border of the shaded area.
\citet{NNK+06} reported a higher value of $\beta=1.99$ which is not
compatible with our observations.
The \citet{WD01} (hereafter WD01) model is overlaid for comparison. The
variations in the extinction law found in this work are rather small compared
to the dust evolution from $R_V=3.1$ to $R_V=5.5$. 
The variation being observed at visual extinction larger than 20~mag
is likely the main reason for this dust transition to be unknown.
In case A the extinction parameters at 4.5 and 5.8~$\mu$m are smaller
at large extinction than at small extinction which is not realistic
according the WD01 model. This case teaches us the small deviation
observed in Fig.~\ref{f.HK-KI1} cannot be ignored as it would make 
longer wavelength results inconsistent.
Case B is more likely even if it yields a larger value than reported in
the literature until now. The largest value prior to this work was from
\citet{CMLE09} with $A_{[3.6]}/A_{K_s}=0.64$. 
However their numbers are overestimated compared to ours or to 
\citet{FPM+07} because they chose a smaller $\beta$.
Their results come from an average for Ophiuchus, Perseus and the Serpens
molecular clouds which all form low-mass stars whereas the cloud associated
with the Trifid is clearly more massive and forms OB stars. \citet{FPM+07}
investigate the high-mass star-forming region of Orion and obtain a 
similar number of $0.636\pm0.003$, significantly below our 
$A_{[3.6]}/A_{K_s}=0.678\pm0.006$. They did find smaller values at all
wavelengths in low-mass star-forming regions.
The presence of massive star, in addition to a high density, may therefore
be relevant.

WD01 attribute the variation of the extinction law with $R_V$ to
grain growth only, although ice mantles onto dust grains contribute to
the optical property variations. The main ice features relevant to our
study originate from H$_2$O and CO$_2$ ices with absorption at 3.0~$\mu$m
and 4.27~$\mu$m, respectively. 
Both H$_2$O and CO$_2$ ices are detected at visual extinction below 4~mag
\citep{WGHS01,BMG+05} where the WD01 model is still valid. This indicates
ice is not sufficient to significantly modify the extinction law in the
Spitzer/IRAC filters. We can also rule out the possibility that ice features
saturate at $A_V\approx20$~mag because they are observed at larger
extinction without any saturation (A. Boogert, private communication).
An example of spectrum is presented in \citet[Fig.1]{WGT+98} for a star with
$A_V=21$~mag. It worth noting the H$_2$O ice band at 3.0~$\mu$m peaks 
outside the [3.6] IRAC filter (3.2-3.9 $\mu$m).
Beside the absorption features, ice mantles make the grains more sticky,
helping the coagulation process. Using Spitzer/IRS spectra \citet{Mcc09}
found evidence for a flattening of the extinction law at $A_K>1$~mag which
would be due to the coagulation of ice-mantled grains.
This suggests ice makes the coagulation process more efficient but it is
not the dominant parameter since the coagulation starts well after
ice-mantles appear. 
The abrupt change in the extinction law at $A_V\approx 20$ is not
understood and we can only speculate that the high density in the Trifid
cloud might allow a new structuring of the dust grains or that the shocks
observed in the Trifid cloud could be determinant.
In a recent work \citet{Boo11} also reported a flatter extinction law
than WD01 based on a spectrum from 3 to 24~$\mu$m for a star with 25~mag
of visual extinction.

\section{From color to extinction}\label{s.extmap}
\subsection{Composite map}
\begin{figure*}
\includegraphics[width=18.0cm]{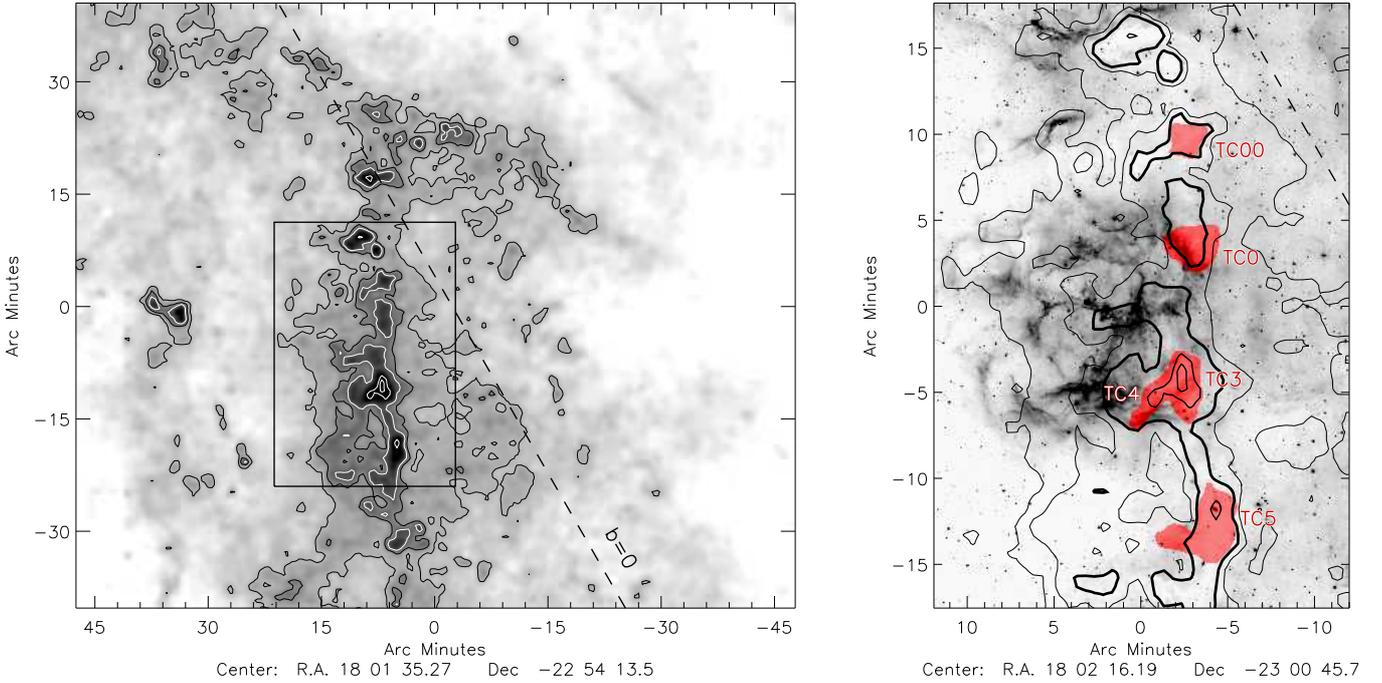}
\caption{{\em Left:} large scale extinction map of the Trifid region. 
{\em Right:} GLIMPSE 8~$\mu$m image of the Trifid nebula with iso-extinction
contours. In both figures contours correspond to $A_V=[10,15,20,40,60]$~mag,
the dashed line follows the Galactic plane. The red shaded regions delineate
the $S_{1.25}>10$~mJy/11''~beam Trifid Condensations \citep[TC,][]{CLC+98}
studied in Sect.~\ref{s.mass} at a 1~arcmin resolution.}
\label{f.avtrifid}
\end{figure*}
The extinction law parameters derived in the previous section allow us to
convert the color maps into extinction. To proceed we follow the more
realistic case B for $A_\lambda/A_{K_s}$ when $A_V>20$~mag (see
Table~\ref{t.res1}) and a reference color of $H-K_s=0.54$~mag.
The associated zero points derived for the other colors are $K_s-[3.6]=0.497$,
$[3.6]-[4.5]=-0.017$ and $[4.5]-[5.8]=0.187$~mag. The sensitivity of each map
relies on the wavelengths used to build the color maps. Shorter wavelengths
are more sensitive too smaller extinctions. The final extinction map
presented in Fig.~\ref{f.avtrifid} is a combination defined as follow:
$$
A_V \mbox{ from } \left\{
\begin{array}{crll}
&K_s-[3.6]   &\mbox{ if } & A_V<15 \mbox{ mag}\\
&[3.6]-[4.5] &\mbox{ if } & 10<A_V<80 \mbox{ mag}\\
&[4.5]-[5.8] &\mbox{ if } & A_V>60 \mbox{ mag}
\end{array}
\right.
$$
The transition between two maps is progressive and linear within the
overlapping extinction ranges. A maximum visual extinction of $\sim$80~mag
is reached. In complement to the extinction map Fig.~\ref{f.avtrifid} shows a
close-up on the Trifid nebula observed at 8~$\mu$m with the
iso-extinction contours superimposed. The parent molecular cloud is an
elongated structure with its densest region lying just West of the
famous optical nebula.

\subsection{Tridimensional analysis, distance estimation}\label{s.3d}
The extinction estimation relies on background stars. The meaning of {\em
background} is not that trivial in the direction of the Trifid nebula since
the line-of-sight crosses the whole galactic disk at only 1~kpc of the
Galactic center. The complexity of the line-of-sight and the distance
of the stars used for the analysis are critical and can change our
interpretation of the color maps. Even the distance of the nebula is
still uncertain. To address these issues we performed a tridimensional
analysis which aims to separate the different components along the
line-of-sight. 

\begin{figure}
\includegraphics[width=8.8cm]{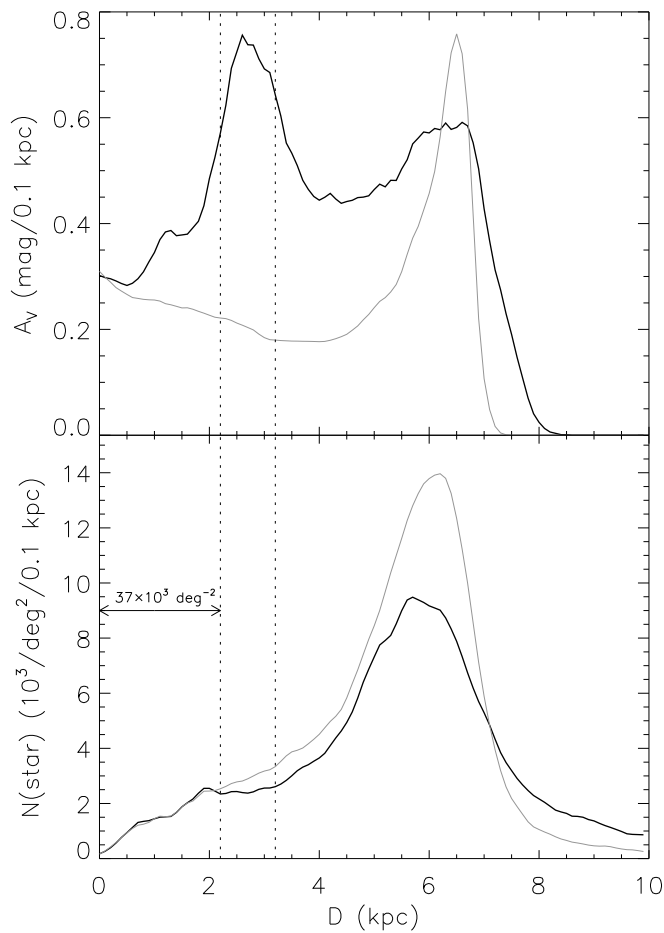}
\caption{Extinction (top) and star density (bottom) along the Trifid
line-of-sight. The black line is the average value for the region
where $A_V>20$~mag in Fig.~\ref{f.avtrifid}; the gray line is for the
region where $A_V \in [0,1]$~mag. The dashed lines indicate the
Trifid location at a distance of $2.7\pm0.5$~kpc.}
\label{f.los}
\end{figure}

A three dimensional stellar model of the Galaxy \citep{RRDS03} is used
to simulate the stellar distribution along the line-of-sight towards the
Trifid nebula. The simulated stars are not subjected to extinction and so
they provide an estimate of the intrinsic color and distance of the stars
along this line-of-sight. By comparing the color distribution of the
unreddened, simulated, stars to the observed stellar color distribution
it is possible to infer the extinction distribution along the 
line-of-sight \citep{MRR+06}. 
We use observations from 2MASS, UKIDSS and GLIMPSE and use the Ks-[3.6]
color to probe the line-of-sight extinction.
This technique is relatively independent of the stellar counts and reflects
the changing stellar color as a function of distance (see \citealt{MRR+06}
for more details). Furthermore there is no need to remove foreground stars
as we are comparing the observations to a model where all stars are included.
However, care needs to be taken to ensure that the model is indeed
representative of the observations, so the simulated catalog is cut
at the completeness magnitude of the observations in the $K_s$, [3.6]
and [4.5] bands.
Recently, \citet{MJJ09} improved on this method using a genetic
algorithm to deduce the line-of-sight extinction necessary to reproduce
the observed stellar color distribution along a given line-of-sight. This
version is the one we chose to use as it provides higher angular resolution
($\sim$3~arcmin compared to 15~arcmin previously) and more sensitivity to
extinction features within 2~kpc from the Sun.
The resulting distances are subject to an uncertainty of roughly 15\% but
this may be higher at large distances and through high column densities.
This results in a smearing out of extinction features along the line of
sight, and could explain why, in Fig.~\ref{f.los}, a narrow extinction
peak is found at 6.5~kpc when the extinction is low but this same peak
is found to be wider through the higher density lines-of-sight crossing
the Trifid nebula. At such a distance the interstellar material can be related
to several Galactic structures which are the molecular ring (about 3-4~kpc
from the Galactic center), the start of the Norma Arm, and the
extremity of the Galactic bar.

The results of the 3D modeling is presented in Fig.~\ref{f.los}.
The extinction profile for the region where $A_V>20$~mag (see
Fig.~\ref{f.avtrifid}) provides us with a direct estimation of
the cloud distance. Fitting a Gaussian yields a distance of
$2.7\pm0.5$~kpc. The extension of the cloud along the line-of-sight
is likely from 50 to 100~pc so the Gaussian standard deviation
represents the uncertainty inherent to the 3D modeling, not the cloud size.
The literature often refers to 1.7~kpc \citep{LCO85} for the Trifid nebula
distance although a serious alternative is proposed by \citet{KML99} with
a distance between 2.5 and 2.8~kpc. Both estimations are based on the
photometry of the central stars. The second work is more complete as
it uses more stars and includes spectroscopy and accurate photometry.
Our analysis of the tridimensional stellar distribution unambiguously
supports the long distance hypothesis which place the Trifid in the
Scutum Arm \citep{Val08} rather than the Sagittarius Arm as generally
reported. In the following we assume $d=2.7$~kpc.

For such a distance, the foreground star number is expected to be at
least $37\times10^3$~deg$^{-2}$ (see Fig.~\ref{f.los}), or
$11\times10^3$~deg$^{-2}$ if we restrict the sample to 2MASS sources.
This is in good agreement with the analysis presented in
Sect.~\ref{s.selection} where we found a total foreground star density
of at least $37\times10^3$~deg$^{-2}$ towards the dense region of the
cloud and in agreement with the distribution of 8600~stars~deg$^{-2}$
detected by 2MASS which was found to be uncorrelated with the extinction,
i.e. foreground to the cloud. It worth noting the Sect.~\ref{s.selection}
results are also incompatible with the short distance hypothesis. 
The $^{12}$CO velocity obtained from \citet{DHT01} shows a velocity
structure consistent with our extinction distribution which
corroborates our estimation. The Trifid molecular clouds LSR
velocity is about 18-20~km~s$^{-1}$. Assuming a Sun kinematic from
\citet{MB10}, a distance to the Galactic center of 7.8~kpc and a
rotation speed of 247~km~s$^{-1}$ the Trifid velocity corresponds
to a kinematic distance of 3.2-3.4~kpc which is even larger than our
estimation based on the stellar population model.

The 3D analysis estimates the diffuse extinction in this direction to
$\sim$3~mag~kpc$^{-1}$ and it shows that no other dense cloud significantly
contaminates the Trifid line-of-sight.
The stellar fraction at distance larger 8~kpc represents only 7.7\% of the
background stars at large extinction and 2.6\% at small extinction. In both
cases their median distance is $\sim$5.8~kpc.
The zero point chosen for the extinction map (Fig.~\ref{f.avtrifid}) is
validated by the 3D study as 1) there is no peak at $\sim$3~kpc at low
extinction and 2) it compensates for the diffuse extinction on the
line-of-sight and the contamination at 6.5~kpc.

\subsection{Cloud mass}\label{s.mass}
\begin{figure}
\includegraphics[width=8.8cm]{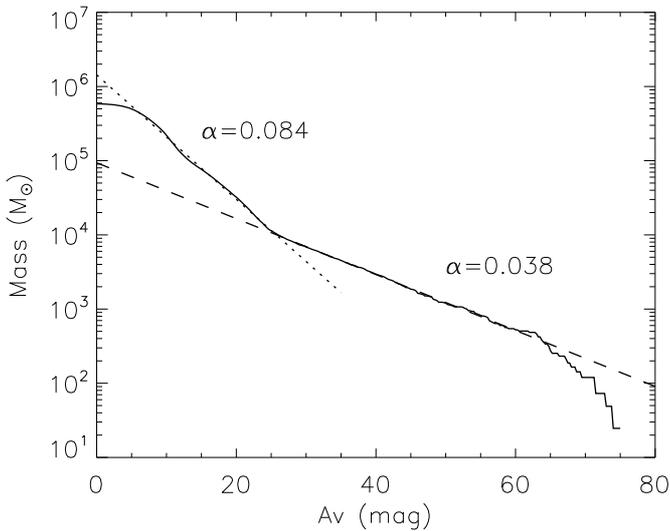}
\caption{Cumulative mass distribution for d=2.7~kpc. The $\alpha$ index
is defined in Eq.~\ref{eq.beta}.}
\label{f.mass}
\end{figure}

The cloud mass is derived from the extinction map using the following
relation \citep{Dic78}:
\begin{equation}
M = (\alpha d)^2 \, \mu \, \frac{N_H}{A_V} \sum_i A_V(i)
\end{equation}
where $\alpha$ is the angular size of a pixel map, $d$ the distance to the
cloud, $\mu$ the mean molecular weight corrected for the helium abundance,
and $i$ represents a pixel map. With the dust-to-gas ratio proposed by
\citet{SM79}, $N_H/A_V=1.87\times 10^{21}$~cm$^{-2}$~mag$^{-1}$
($N_H=N_{HI}+N_{H_2}$) and a distance of 2.7~kpc, we obtain a mass of 
$5.8 \times 10^5$~M$_\odot$. This value corresponds to twice the Orion
(A+B) clouds mass \citep{Cam99d,WDMT05}.
Figure~\ref{f.mass} shows how the integrated mass distribution varies with
the extinction within the cloud. Regions with visual extinction greater
than 20~mag account for less than 4\% of the total mass of the cloud.
The index of the cumulative mass distribution, $\alpha$, changes at
$A_V \approx 25$~mag by a factor of 2. For $M(A_V)$ being the mass
enclosed in the contour of extinction $A_V$, we have:
\begin{eqnarray}
\lefteqn{ M(A_V) = M(0) \times 10^{-\alpha A_V} } \label{eq.beta}\\
&&\mbox{ with } \left\{
\begin{array}{l}
\alpha = 8.39\times 10^{-2}  \mbox { for } A_V \in [8,20]\\
\alpha = 3.76\times 10^{-2}  \mbox { for } A_V \in [30,60]
\end{array}
\right.\nonumber
\end{eqnarray}

The variation of $\alpha$ is not a side-effect of the map combination
technique nor the application of a corrected extinction law at large $A_V$
since the variation is also observed for a raw extinction map obtained
only from the [3.6]-[4.5] color-excess and assuming a constant extinction law.
It is however difficult to interpret the index variation as a consequence of
the dust transition responsible of the extinction law evolution. 
The two events occurring at the same level may just be a coincidence. 
The $N_H/A_V$ value is generally assumed to be universal although some
authors suggested a decrease by a factor of 3 in deeply embedded regions
\citep{WMW+10}. Dividing the mass estimate by any factor above
a given extinction threshold would only produce an offset in
Fig.~\ref{f.mass}, not a change of slope. A more complex
evolution of $N_H/A_V$ with the extinction would affect differently
the Fig.~\ref{f.mass} distribution but it seems unlikely to only produce
a sharp slope change.

\begin{table*}
\begin{minipage}[t]{18cm}
\caption{Properties of the more extended Trifid cores detected at millimeter
wavelength \citep{LCP08}. Masses are expressed assuming a distance of
2.7~kpc. The mass from dust continuum emission is derived using
$T_{\rm dust}=15$~K and $\kappa_{1.25} =4.08\times10^{-1}$~cm$^2$~gr$^{-1}$ 
(i.e. $C_{1.25}=7.336\times10^{-27}$~cm$^2$~H$^{-1}$).
By modifying these parameters the mass can be adjusted to the mass
estimated from the dust extinction. The two last columns give these 
parameter values, providing only one of the two parameters is modified at the
same time.}
\label{t.TC}
\renewcommand{\footnoterule}{}  
\centering
\begin{tabular}{lcccccc|cc}
\hline\hline
Name & Mass\footnote{Mass from the dust near-infrared extinction.} & 
Mass\footnote{Mass from the dust continuum millimeter emission.}   &
\multicolumn{3}{c}{$A_V$ (mag)} & Surface & $\kappa$ & $T_{\rm dust}$ \\
$[$CLC98$]$ & ($M_\odot$) & ($M_\odot$) & peak & median & outer &
(arcmin$^2$) & ($\times$)& (K)\\
\hline
TC00 &  936 &  911        & 27         & 22.7 & 20.0  & 3.5  & 1.3 & 13 \\
TC0  & 1607 & 1570        & 25         & 20.3 & 16.9  & 6.9  & 1.2 & 13 \\
TC3  & 3535 & 2783        & 78         & 40.4 & 27.6  & 7.3  & 2.6 &  8 \\
TC4  & 1149 &  945        & 52         & 32.8 & 23.7  & 3.0  & 2.7 &  8 \\
TC5  & 3543 & 3087        & 53         & 25.5 & 18.1  & 11.6 & 1.7 & 11 \\
\hline
\end{tabular}
\end{minipage}
\end{table*}

The millimeter wavelength emission from the cold dust continuum has been
mapped by \citet{CLC+98} and \cite{LCP08}. They have detected several dust
condensations from the 1.25~mm emission. For simplicity we designate
these dust condensations TC~NN as in the original papers. This is fine is
the Trifid context but obviously ambiguous in the general context and the
full SIMBAD name is actually [CLC98]~TC~NN. 
For the more extended condensations we compare the mass from the dust
infrared extinction and 1.25~mm emission. The results are presented in
Table~\ref{t.TC}.
The mass from extinction is obtained by integrating the area defined in the 
map at 1.25~mm, convolved at 1~arcmin resolution, by the contour at
$S_{1.25}=10$~mJy$/11''$~beam (red shaded area in Fig.~\ref{f.avtrifid}).
Because of the observational technique in the millimeter wavelengths, a
spatial filtering removes the structure at scale larger than 4.5~arcmin
in the dust continuum emission. To compare this map with the extinction
map we scale it by adding the median extinction measured in the vicinity of
each TC region.
Basically the median extinction is measured in the area between the
contours at $S_{1.25}=5$ and 10 mJy/11'' around each condensation. The
resulting offsets are given in the column labeled {\em outer} of
Table~\ref{t.TC}. 
The mass from the map at 1.25~mm is derived assuming a dust temperature of
15~K and a dust opacity of $\kappa_{1.25}=4.08\times10^{-1}$~cm$^2$~gr$^{-1}$
following the model B for $R_V=5.5$ by WD01. 
The corresponding extinction cross section per hydrogen atom is 
$C_{1.25}=7.336\times10^{-27}$~cm$^2$~H$^{-1}$ after correction for the
helium abundance and assuming a gas-to-dust ratio of 125 \citep{LD01}.
\citet{LCP08} measured temperatures from 20 to 30~K using CO observations.
The gas kinetic temperature and the dust temperature are however different
in such regions and an arbitrary $T_{\rm dust}=15$~K is more appropriate.
The masses we derived from the dust emission are systematically smaller.
Two parameters, the temperature and the emissivity, can be adjusted to
reach an agreement with the masses from extinction.
While TC00 and T0 only need a small increase of the dust emissivity at
1.25~mm, or to decrease the temperature from 15 to 13~K, TC3/4/5 require a
more significant emissivity increase by a factor 1.7 to 2.7. Using only the
temperature as free parameter would lead to 11 to 8~K. Such low temperatures 
have already been reported by \citet{DMM+08} from submillimeter point
sources observed by the Archeops experiment. However, the condensations we
are dealing with cover an extended field from 3 to 12~arcmin$^2$.
Assuming a minimum temperature of 12~K for TC3/4, an emissivity
enhancement by a factor 1.9 would make the mass estimates in agreement. 
TC00 and TC0 are starless condensations while TC3/4/5 form stars and
undergo shocks. Our analysis suggest the dust emissivity at millimeter
wavelength is enhanced for these condensations, especially for the densest
ones. 
Complementary observations at longer wavelengths, from Herschel and Planck,
will permit to better understand the physical properties of the interstellar
medium in the Trifid region. We plan to follow up this study as soon as
the data from the {\em Herschel Infrared Galactic Plane Survey}
\citep[Hi-GAL,][]{MSB+10b} are available.

\section{Conclusion}\label{s.conclusion}
We performed a detailed analysis of the extinction in the molecular cloud
associated with the Trifid nebula. The direction and the distance of this
cloud is a major difficulty compared to the other similar investigations
published in the literature. To overcome this specificity we proposed an
original method which combines the color-excess mapping with the color-color
diagram analysis. The gain in sensitivity with this method allowed us to
measure to extinction law at 3.6, 4.5 and 5.8~$\mu$m and to unambiguously
detect a variation at large absorption through an abrupt change of slope
in the color-color plots. We interpreted this result as an evidence for
a rapid dust evolution at high density like a new dust transition phase
in the dense interstellar medium.

The extinction law parameters, $A_{\lambda}/A_{K_s}$, are found to be larger
in the very dense cores of the cloud, in agreement with dust models. Our
values for $A_V>20$~mag are however larger than predicted by WD01
for dense regions ($R_V=5.5$) and larger than previously reported in the
literature at 3.6~$\mu$m. For $A_V<15$~mag, which does still refer to dense
material, our values remarkably match those predicted by the model for
$R_V=5.5$, as expected. Our study is not sensitive to the diffuse medium
for which $R_V=3.1$ with a typical visual extinction as low as 1 mag.
Moreover, \citet{ZMI+09} found a correlation of the extinction law
for the diffuse interstellar medium with the galactocentric radius. They
concluded the extinction curve follows the dust model with $R_V=5.5$ for
a galactocentric radius consistent with the Trifid distance. It suggests
the so-called diffuse medium already contains big grains in this part of
the Galaxy. If the diffuse medium at the Trifid distance behaves like the
dense medium in the solar neighborhood we can imagine the Trifid dense
cores are also affected by the distance with a flatter extinction law.

Using our varying extinction law we built a composite extinction map of the
Trifid nebula region. The maximum visual extinction is 80~mag at 1~arcmin
resolution. We have also investigated the matter distribution along the
line-of-sight since several small clouds could have mimicked the presence
of a single giant massive dark cloud. We found no significant contribution
of other clouds in this direction and we were able to estimate the distance
of the Trifid nebula at $2.7\pm0.5$~kpc.

The Trifid molecular cloud is about twice the mass of the Orion molecular
cloud. The comparison of the dust extinction and 1.25~mm continuum emission
in the densest cores suggests the dust emissivity is probably enhanced by a
factor of 2-3 in these regions. 
This cloud appears as a young and massive version of the Orion or Rosette
regions. The Trifid represents an earlier evolutionary status in the
star-formation process  with a first generation of OB stars but no
significant embedded cluster yet. Several protostars in TC3/4 are likely
the precursors of such clusters. 
Longer wavelengths observations from Herschel and Planck are needed to
better understand the dust properties and the interaction between the
interstellar medium and the OB stars.

\begin{acknowledgements}
We thank the anonymous referee for his comments that help to clarify and 
improve this paper.
We thank T. Dame for providing us with the CO datacube and for his
assistance in using it. We are also grateful to A. Boogert for his help
understanding the possible ice contributions to our analysis.
This work is based in part on observations made with the Spitzer Space
Telescope, which is operated by the Jet Propulsion Laboratory, California
Institute of Technology under a contract with NASA.
This work is based in part on data obtained as part of the UKIRT Infrared
Deep Sky Survey.
This publication makes use of data products from the Two Micron All Sky
Survey, which is a joint project of the University of Massachusetts and
the Infrared Processing and Analysis Center/California Institute of
Technology, funded by the National Aeronautics and Space Administration
and the National Science Foundation
\end{acknowledgements}

\bibliographystyle{aa}
\bibliography{biblio}

\end{document}